\documentstyle[12pt]{article}
\input epsf
\begin{document}
\baselineskip 15pt
\rightline{CU-TP-657} \vskip 2cm
\centerline{\large\bf Magnetic Black Hole Pair-Production:}
\centerline{\large\bf One-Loop Tunneling Rate in the Weak Field
Limit}

\vskip 1.5cm
\centerline{{ \sc Piljin Yi}\footnote{e-mail address:
piljin@cuphyf.phys.columbia.edu}}
\vskip 3mm
\centerline{\it Physics Department, Columbia
University, New York, NY 10027, USA}
\vskip 3cm
\centerline{\bf Abstract}
\vskip 3mm
\begin{quote}
\noindent {\baselineskip 12pt {\small
Pair-production of magnetic Reissner-Nordstr\"{o}m black holes (of charges
$\pm q$) is studied in the next-to-leading WKB approximation. We consider
generic quantum fluctuations in the corresponding instanton geometry, a
detailed study of which suggests that, for sufficiently weak field $B$,
the problem can be reduced to that of quantum fluctuations around a single
truncated near-extremal Euclidean black hole in thermal equilibrium.
Typical one-loop contributions are such that the leading WKB exponent is
corrected by a small fraction $\sim \hbar /q^2$. We show that this effect
is merely due to a semiclassical shift of the black hole mass-to-charge
ratio that persists even in the extremal limit. We close with a few final
comments.}}
\end{quote}

\vskip 2cm
\centerline{\it A talk presented in the $XIII^{th}$
Sorak International Symposium}
\centerline{\it 'Field Theory and Mathematical Physics,'
June 1994.}

\newpage

Recently, the pair-production of oppositely charged magnetic black holes
in a background magnetic field, has been studied to the leading WKB
approximation \cite{ERNST}\cite{GGS}\cite{GAUNT}\cite{GG}.
The instanton mediating the tunneling process is found to be the
Euclidean section of the so-called Ernst metric, and the Euclidean
action thereof has been calculated exactly.

One cannot emphasize too much the importance of such  processes in the
context of black hole quantum physics. The apparent conflict between the
unitarity principle and the black hole evaporation \cite{Info}
spurred many different speculations \cite{remnants}
regarding true nature of the black hole, most of which
cannot be tested from the usual semiclassical reasonings leading to
the presence of Hawking radiation. Spontaneous pair-productions of
charged black holes, then, provides an alternative theoretical laboratory
where one may gain further insight on this controversial issue.

As a part of effort in this regard, we want to study the tunneling event
when the pair-produced objects are near-extremal Reissner-Nordstr\"{o}m
black holes. One interesting property of the extremal limit is that the
proper distance to the horizon diverges, suggesting very large spatial
volume associated with the black hole, that might contain many degenerate
states \cite{Trivedi2}.
Our main purpose here is to determine whether this leads to
a huge enhancement of the tunneling rate at one-loop level, or even a
breakdown of the semiclassical approximation.  More complete account of
the present work appeared elsewhere \cite{WKB}.

\vskip 5mm

Let us first write down the instanton metric that mediates the
pair-production of arbitrary magnetic Reissner-Nordstr\"{o}m
black holes\cite{ERNST}\cite{GAUNT}.
\vskip 1mm
\begin{eqnarray}
g^{(4)}&=&\frac{\Lambda^2}{(Ay-Ax)^2}\bigl(-G(y)dT^2-\frac{dy^2}{G(y)}\bigr)
\nonumber \\
       &+&\frac{1}{(Ay-Ax)^2}\bigl(\frac{\Lambda^2}{G(x)}dx^2
                +\frac{G(x)}{\Lambda^2}d\phi^2\bigr) \label{eq:E-E} \\
G(\xi)&=&(1+r_{-}A\xi)(1-\xi^2-r_{+}A\xi^3) \\
\Lambda&=&\Lambda(x,y)=(1+\frac{1}{2} qBx)^2+\frac{B^2}{4(Ay-Ax)^2}G(x)
\end{eqnarray}
\vskip 5mm
\noindent
The geometry comes with two Killing coordinates $T$ and $\phi$, where the
latter generates an axial symmetry. The Minkowskian version of this metric
describes the geometry after the materialization of the black holes which
are subsequently accelerated by the Lorentz force. For sufficiently
small external field in particular, $B$ and $A$ correspond to the magnetic
field strength on the symmetric axis and the magnitude of acceleration.
The remaining three parameters are related by $q\equiv
\sqrt{r_+ r_-}$ and $0 <r_- \le r_+$.

The audience may find this metric a little bit disturbing, for it
appears completely static with the time coordinate being $T$.
How do we know that the black holes are really accelerating?
The resolution of the puzzle is easily obtained by studying
the geometry far away from the pair-created objects.
After a couple of coordinate redefinitions \cite{GG}:
\begin{equation}
\zeta^2= \frac{y^2-1}{A^2(x-y)^2},\qquad \rho^2=\frac{1-x^2}{A^2(x-y)^2}.
\label{eq:New}
\end{equation}
the metric can be rewritten as follows provided that $-r_\pm Ay\ll 1$,
\begin{equation}
g^{(4)}\simeq \Lambda^2\,(  \zeta^2\,dT^2+d\zeta^2+d\rho^2)+\Lambda^{-2}\,
\rho^2\,d\phi^2,\qquad \Lambda\simeq 1+\frac{B^2\rho^2}{4}. \label{eq:Rindler}
\end{equation}
This constitutes a background space containing an external magnetic field,
referred  to as a  Melvin universe.
Note that, up to the warping influence of $B$, the metric looks like
that of Rindler where a family of accelerating observers appear static
with respect to their common proper time. Therefore, an object static
with respect to the Killing coordinate $T$ is actually experiencing an
acceleration roughly given by $1/\zeta\Lambda$. In particular, it turns out
that the black holes, if small compared to $B^{-1}$, are at $\zeta\simeq
1/A$ and $\rho \ll 1/A$, partially  confirming our statement that the black
holes are accelerating at $A$. Furthermore, as is well-known among
researchers, such a uniformly accelerating object behaves as if it is
in a heat bath of temperature $T_{Unruh}\simeq \hbar A/2\pi$ \cite{Unruh}.
This fact will be of central importance in studying the tunneling process
later on.

\vskip 5mm
Are the pair-produced black holes really of Reissner-Nordstr\"{o}m
type? In order to separate out distortions due to the accelerating
motion of these objects, it is sufficient to consider the case $r_+ A
\rightarrow 0$. Expanding in terms of a new radial coordinate $rA\equiv
-1/y \ll 1$,
\begin{eqnarray}
(Ay-Ax)^2&=&\frac{1}{r^2}+\cdots,\\
G(y)&=&-\frac{1}{r^2A^2}(1-\frac{r_{-}}{r})(1-\frac{r_{+}}{r})+\cdots,  \\
G(x)&=&1-x^2+\cdots, \\
\Lambda&=&1+\cdots.
\end{eqnarray}
and introducing two more coordinates $\tau\equiv T/A$ a Euclidean time,
and $\theta=\cos^{-1} x$ the azimuthal angle \cite{GAUNT},
we find an approximate form of the geometry near the black hole valid
for small acceleration:
\begin{equation}
g^{(4)}\simeq (1-\frac{r_{-}}{r})(1-\frac{r_{+}}{r})\,d\tau^2+
         \frac{1}{(1-\frac{r_{-}}{r})(1-\frac{r_{+}}{r})}\,dr^2+
         r^2\,(d\theta^2+\sin^2\theta\,  d\phi^2). \label{eq:R-N}
\end{equation}
This is of course the celebrated Reissner-Nordstr\"{o}m black hole,
whose mass $m\equiv (r_+ +r_-)/2$ is bounded below by its absolute charge
$q=\sqrt{r_+ r_-}$. The extremal limit, $r_= \rightarrow r_-$,
saturates this inequality.

\vskip 5mm
In the above limit of small $A$, it is immediately clear that
the configuration described by Eq. \ref{eq:E-E} cannot be a classical
solution for all values of $A$, $B$, and $r_\pm$. For one thing,
to  the leading order, the only force acting on the black holes is the
Lorentz force. This demands the acceleration be proportional to the
external magnetic field:
\begin{equation}
mA\simeq qB. \label{eq:NT}
\end{equation}
Under more general circumstances, in fact, the corresponding constraint
appears from the condition that there is no cosmic strings attached to
the black holes \cite{ERNST}\cite{GAUNT}: a rather fancy way of
deriving the black hole equation of motion.

Ordinarily this would be end of the story: the instanton would be
parametrized by three independent parameters  which are the mass $m$,
the charge $q$ and the external  magnetic field $B$. Changing $m$ with
the other two fixed would simply result in different WKB exponent.
However, unlike point-particles or nonsingular solitons, a black hole
of given charge and mass is not a semiclassically stable object in
general. Whenever its Hawking temperature is nonzero, it radiates
thermal radiations that tends to decrease its mass steadily in the
absence of any matter influx \cite{Hawking}:
\begin{equation}
-\dot{m} > 0
\end{equation}
This poses a conceptual difficulty in gluing the imaginary-time
description of the tunneling to the real-time evolution thereafter,
since such a transition is possible only when the configuration is
truely free from any time dependence at the moment.
For this reason, a semiclassically consistent calculation requires
the black holes to be in thermal equilibrium with their environment
when they materialize:
\begin{equation}
T_{BH}\simeq\frac{\hbar\, (r_+ -r_-)}{4\pi r^2_+}\: \simeq \:
\frac{\hbar A}{2\pi} \simeq T_{Unruh}. \label{eq:TEM}
\end{equation}
The accelerating black hole not only radiates at temperature $T_{BH}$ but
also accretes thermal quanta from the surrounding Rindler heat bath
\cite{Unruh} alluded earlier. The equal temperature is therefore
necessary to ensure time-independence at the moment of materialization.

This requirement turns out to be encoded naturally in the singularity-free
geometry of the Euclidean instanton \cite{ERNST}\cite{GAUNT},
in much the same fashion as a
singularity-free Euclidean black hole knows about its Hawking temperature.
Together with the Newton's law (Eq. \ref{eq:NT}), this fundamental
constraint dictates that the weak field limit is identical to the extremal
limit $r_+ \rightarrow r_-$, thus simplifying our task considerably.

\vskip 5mm
In the same weak field limit, the leading WKB exponent, as determined by
the Euclidean action of the instanton \cite{ERNST}\cite{GGS}\cite{GG},
can be expanded as follows,
\begin{equation}
-\frac{S_{E}}{\hbar}=-\frac{\pi q}{\hbar B}+\cdots=-\frac{\pi m^2}{
\hbar qB}+\cdots \label{eq:Schwinger}
\end{equation}
The last expression is easily recognized as the Schwinger term that also
appears in the monopole pair-production \cite{MONO}, while the ellipsis
denotes terms of higher order in $qB$, and includes the Bekenstein-Hawking
entropy $+{\cal S}_{BH}\simeq \pi q^2$ \cite{BEK}.

\vskip 5mm
Now before we ask how strong is the one-loop correction to this as $qB\simeq
r_+A\rightarrow 0$, it is instructive to see the overall shape of the
instanton geometry. We have already seen that, for small values of $qB
\simeq r_+A $, the geometry is approximated by simpler ones near and far
away from the black hole, More precisely, the geometry is that of a single
Eucludean black hole for $-1/y \ll 1$ while it is that of a Melvin universe
for $-r_+ A y \ll 1$. In addition, the two approximate regions actually
overlap with each other for even smaller $r_+ A$: The full instanton
consists of these two simple geometric components connected via a
transtional region where both approximations make sense.
\begin{equation}
1 \ll -y \ll \frac{1}{r_+ A}
\end{equation}
In terms of the Rindler-like coordinates of the background Melvin space,
this corresponds to $\zeta\simeq 1/A$ and $\rho \ll 1/A$ where a Euclidean
black hole {\it without the asymptotic region} must be attached smoothly to
the background space \cite{WKB}.
To see where the transition occurs in term of $r$, it is sufficient
to compare how typical curvature scales for each component: curvatures of
the black hole geometry scales like $q/r^3$ while, for small $\rho$, the
Melvin space is curved at a typical scale of $B^2$. Equating the two, we
find that the black hole must be truncated at $r=r_B\sim (q/B^2)^{1/3}$
\cite{WKB}. This simple picture of the instanton is depicted in figure 1.
without the correct scale.
\vskip 15mm
\begin{center}
\leavevmode
\epsfysize=1.5in \epsfbox{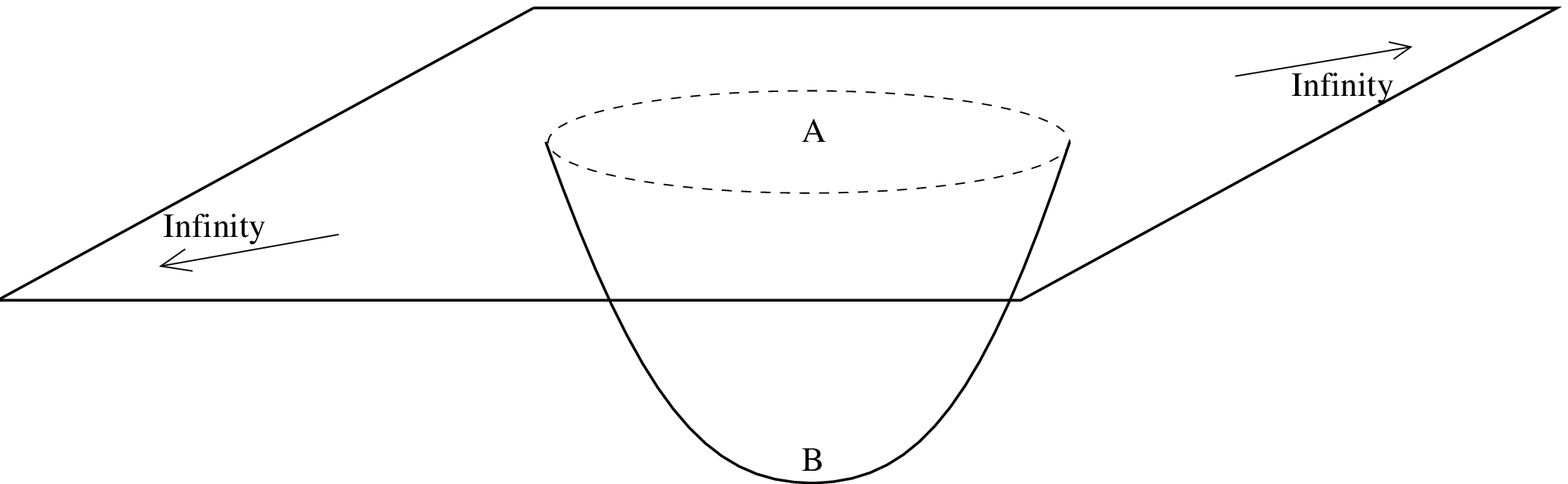}
\end{center}
\vskip 5mm
\begin{quote}
{\bf Figure 1:} {\small A schematic diagram for the Euclidean instantons.
In the weak field limit, the bottom ``cup'' is described by a near-extremal
Euclidean black hole, while the top ``sheet'' is the Euclidean Melvin space.
The acceleration horizon and the black hole horizon are located at points
A and B, respectively. The transitional ``mouth'' region is denoted by the
broken curve where the area of the transverse two-sphere is $\sim 4\pi\,
q^{2/3} B^{-4/3}$. }
\end{quote}
\vskip 10mm

An interesting property of this geometry is the divergent 4-volume of the
bottom ``cup'' region, or the truncated Euclidean black hole,  as $qB
\rightarrow 0$: divergent not only because of the increasing circumference
along the time direction but also because of the logarithmically divergent
distance to the black hole horizon. On the other hand, if we were studying
pair-production of ordinary charged particles in flat spacetime, the truncated
Euclidean black hole would be replaced by a simple circular trajectory
of the particle in question, and there would be no divergent extra 4-volume
to speak of. This again remind us of the primary question whether the
extremal limit in this context involves huge quantum corrections from many
low-energy modes uniquely associated with black holes.

\vskip 5mm
One of the more important aspect of this Euclidean instanton is the natural
vacuum associated with it. As emphasized above, the nonsingular Euclidean
horizons demand that there exists a thermal equilibrium between the
black holes and their environment, which implies that the vacuum behaves
like a Hartle-Hawking vacuum near the black holes \cite{GG}\cite{WKB}.
But, from a purely Euclidean point of view, Hartle-Hawking vacua are a naive
consequence of the periodic imaginary time, and do not appear particularly
unique to this pair-production process.

In reality, however, the usual physical vacuum around a black hole in real
time cannot be the static Hartle-Hawking vacuum that describes a thermal
equilibrium, but the evolving Unruh vacuum that involves the thermal Hawking
radiation, one of the reason being that the necessary heat bath extends
naturally to the asymptotic region, costing an infinite amount of energy.
This happens because the gravitational redshift settles down to a finite
factor as quanta climes out of the gravitational well. Alternatively,
one simply observes that the proper circumference along the imaginary
time approaches a constant $\hbar/T_{BH}$ asymptotically.

On the other hand, the instanton geometry
does not asymptote to that of a Euclidean
black hole, but rather to that of a Melvin space in Rindler-like coordinates
of Eq. \ref{eq:New}. The asymptotic region is then essentially a flat
spacetime as seen by Rindler observers, up to the gentle warping influence
of magnetic flux.\footnote{This magnetic flux can be cut-off for large
transverse distances from the symmetry axis, for instance, by imagining the
whole process taking place inside a giant cosmic string.} Therefore, the
natural vacuum behaves like a Hartle-Hawking vacuum where $r < r_B$ but
smoothly continues to a Melvin analogue of the ordinary Minkowski vacuum.

The upshot is that the necessary heat baths around the pair-produced black
holes, Euclidean or not, are now of finite size $\sim r_B$. Hence, the
hybrid nature of the instanton geometry makes the semiclassical equilibrium
possible without costing an infinite amount of energy.

\vskip 5mm
Above discussions tell us, among other things, that in a suitable limit
the problem of finding one-loop corrections to the leading WKB estimates may
reduce to a simpler problem involving quantum fluctuations around a single
Euclidean black hole truncated at $r\sim r_B$. It is clear that in the
weak field limit, the classical geometry does split into two regions,
one of which merely provides background fields. Furthermore, the vacuum
state also appears to be ``cut off'' naturally at the boundary between the
two regions. This combined with the fact that we actually calculate the
ratio of two partition functions, one associated with the total instanton
geometry and the other associated  with the background Melvin space,
indeed suggests that we may concentrate on the bottom ``cup'' part of
the instanton \cite{WKB}.

Of course, in principle,
it might be very important exactly how this truncated manifold is attached
to the background Melvin space, but as far as the simple examples we
consider in order to isolate the leading $qB$ terms, this does not appear
to be the case. For the remainder of the talk, we will concentrate
on one particular contribution that arises from the so-called Callan-Rubakov
modes, for which an explicit calculation is available. Estimates
for more general cases as well as a detailed version of the
discussions so far can be found elsewhere \cite{WKB}.

\vskip 5mm

For most quantum fluctuations around the instanton solution, there
exist potential barriers near the black hole event horizon.
For instance, angular momentum $l$ modes of a minimally coupled massless
scalar find the following potential barrier $V_l$ in the tortoise coordinate
$z$.
\begin{equation}
g=F(z)\,(-dt^2+dz^2)+R^2(z)\,d\Omega^2\qquad \Rightarrow \qquad V_l(z)=
\frac{\partial_z^2 R}{R}+\frac{l(l+1)\, F}{R^2}
\end{equation}
Such potential barriers are especially inhibitive for low energy
excitations which are responsible for the infrared behaviour of the
effective action.

One exception to this is the celebrated Callan-Rubakov modes
\cite{Callan}\cite{Holzhey} in spherically symmetric magnetic field
backgrounds, chargeless combinations of which propagate effectively as 2-D
conformal fields \cite{Alford}\cite{theta}. While the instanton geometry
is not spherically symmetric everywhere, we have seen that, in the weak field
limit $qB\rightarrow 0$, the spherical symmetry is restored near the
Euclidean black hole horizon, giving us some hope that the contribution from
these uninhibited modes near the Euclidean black hole may capture the
essential physics of the next-to-leading WKB.

The effective actions obtained by integrating over 2-D conformal fluctuations
are exactly known in an explicitly non-local form, namely the
Polyakov-Liouville action \cite{Polyakov}, up to a local topological
contribution. Accordingly, the contribution to the prefactor  \cite{ColeWKB},
${\cal N}_{\rm CR}$, from $N$ such S-wave modes can be written
as following,
\begin{equation}
{\cal N}_{\rm CR}=
e^{-W}=e^{-N S_{PL}},\qquad S_{PL}=\frac{1}{96\pi}\int dx^2\sqrt{g^{(2)}}\,
R^{(2)}\frac{1}{\nabla^2}R^{(2)}\: + \cdots
\end{equation}
The ellipsis denote a topological term independent of the geometry.
The form of Polyakov-Liouville action
becomes particularly convenient to handle in conformal  coordinates, since the
scalar curvature $R^{(2)}$ above can be expressed simply as a Laplacian
of the conformal mode. Here, let us choose an asymptotically flat
conformal coordinate.
\begin{equation}
g^{(2)}=F\,(d\tau^2+dz^2) \quad \rightarrow \quad R^{(2)}=-\nabla^2 \log F.
\end{equation}

\noindent
For the case of the Euclidean Reissner-Nordstr\"{o}m black hole, in particular,
$F$ and $z$ are given by
\begin{equation}
F=F(r)\equiv (1-\frac{r_{-}}{r})(1-\frac{r_{+}}{r}), \qquad
z=\int\frac{dr}{F(r)}.
\end{equation}

\noindent
The importance of choosing the conformal gauge can be seen from the following
general relationship, where $h$ is any harmonic function on the given
manifold, to be fixed by the boundary condition chosen:
\begin{equation}
\frac{1}{\nabla^2}R^{(2)}=-\frac{1}{\nabla^2}\nabla^2 \log F=-\log F+h .
\label{eq:green}
\end{equation}

\noindent
This way, all the dependence on the  choice of vacuum is encoded into a
single harmonic function $h$. On the other hand, as argued earlier, the black
holes created by the instanton (Eq. \ref{eq:E-E}) are
in Rindler heat-baths of finite size, and the
relevant vacuum must behave like a Hartle-Hawking vacuum near the
black hole horizon. One characteristic of the Hartle-Hawking vacuum is that
the potential divergence of energy-momentum expectation values on both past
and future event horizon disappears.\footnote{In this reagrd, as
emphasized in Ref. \cite{remnants}, choosing
Hartle-Hawking vacuum is not only well-motivated physically but also vital
for the validity of the WKB approximation, for the gravitational
backreaction to the quantum fluctuations is now well controlled. Otherwise,
the quantum fluctuations around the classical solution can no longer be
regarded as ``small'' and a systematic expansion based on the Euclidean
instanton would be ill-fated.}
Moreover, we expect $h$ to be independent of the time coordinate, for
the state is supposed to be in thermal equilibrium. A consistent choice
is then given by
\begin{equation}
h=F'(r_{+})\,z+C \label{eq:HM}
\end{equation}
where $C$ is in arbitrary integration constant.
With this choice, we find {\it semiclassical}
spacetimes  with regular event horizons of non-zero Hawking temperatures
(but with infinite ADM masses due to the heat bath), which is exactly what
one expects from the Hartle-Hawking vacuum \cite{Trivedi}\cite{Jaemo}.

Note that there are some harmless ambiguities remaining. In particular,
an additive shift of the constant $C$ can be translated into an additive
shift of the topological term proportional to the Euler number which
is insensitive to the number $qB$.

Evaluating the Polyakov-Liouville action with $h=F'(r_{+})z+\cdots$,
we find,
\begin{equation}
S_{PL}\biggr\vert_{on-shell} = -\frac{\hbar}{96\pi T_{BH}}
\int_{r_{+}}^{r_B} dr\, \frac{F'(r)\,\bigl(F'(r)-F'(r_{+})\bigl)}{F(r)}
+\cdots. \label{eq:main}
\end{equation}
The radial integral comes with the upper limit at $r_B$, for the
relevant geometry is the truncated Euclidean black hole, but it
matters little owing to the rapidly vanishing behaviour of the integrand.
The resulting integral is finite for any $T_{BH}$, and is continuous
in the extremal limit.
\begin{equation}
\lim_{r_{+}\rightarrow r_{-}}\int^{\infty}_{r_{+}}\frac{F'(r)\,
\bigl(F'(r)-F'(r_{+})\bigl)}{F(r)} =\int^{\infty}_{q}\lim_{r_{+}\rightarrow
r_{-}}\frac{F'F'}{F}=\frac{4}{3q}
\end{equation}

\noindent
Using the constraint $T_{BH}/\hbar \simeq A/2\pi \simeq B/2\pi$,
we arrive at the following one-loop corrected exponent,
\begin{equation}
-\frac{S_{E}}{\hbar}-W= -\frac{\pi q}{\hbar B }+\frac{N}{36qB}
+\cdots=-\frac{\pi q}{\hbar B}\biggl(1-\frac{N\hbar}{36\pi q^2}\biggr)+\cdots .
\label{eq:WKB1}
\end{equation}

\noindent
The ellipsis now denotes terms from quantum fluctuations other than the
chargeless Callan-Rubakov modes as well as terms of higher order in $qB$.

\vskip 5mm
We can also consider more realistic contributions from genuinely
four-dimensional quantum fields. This turns out to yield similar $qB$
dependence, although technical difficulties kept us from
evaluating the relevant coefficients \cite{WKB}.

\vskip 5mm
An immediate consequence of this is that there seems to be nothing
special about the extremal limit other than the fact that the Euclidean
orbit of the magnetic black hole diverges like $1/B$.
In particular, the divergence of $W \sim 1/qB \sim \hbar/qT_{BH}$ in
the extremal limit $T_{BH}\rightarrow 0$ results from the diverging
periodicity $\hbar/T_{BH}$ of the Euclidean time coordinate rather than
from the diverging 3-volume near the black hole horizon. Per unit
Euclidean time, the contribution of quantum fluctuations at one-loop
level remains finite in the extremal limit.

In fact, we believe that this leading one-loop correction can be explained
as a direct consequence of a semiclassically shifted mass-to-charge ratio
of the black holes in question \cite{Jaemo}.
To see this, it is necessary to restore the {\it classical} mass $m\simeq q$
of  these near-extremal black holes:
\begin{eqnarray}
-\frac{S_E}{\hbar}&=&-\frac{\pi m^2}{\hbar qB}+\cdots, \\
-\frac{S_E}{\hbar}-W&=& -\frac{\pi m^2}{\hbar qB}\biggl(1-
\frac{N\hbar}{36\pi q^2}\biggr)+\cdots . \nonumber
\end{eqnarray}
Note that the one-loop corrected expression would look like the original
Schwinger term if we introduce a new mass parameter $\tilde{m}$ as follows:
\begin{equation}
-\frac{S_E}{\hbar}-W=-\frac{\pi \tilde{m}^2}{\hbar q B}+\cdots \quad
\hbox{with}\;\; \tilde{m}\simeq q\,(1-\frac{N\hbar}{72\pi q^2}).
\label{eq:remass}
\end{equation}
This leads us to speculate that this particular one-loop correction
simply represents a semiclassical effect that shifts near-extremal
black hole masses.

If this conjecture is to hold, it is necessary that the mass-to-charge
ratio of extremal Reissner-Nordstr\"{o}m black holes is semiclassically
modified from $1$ to $1-{N\hbar}/{72\pi q^2}$, in the presence of
$N$ chargeless Callan-Rubakov modes. Exactly how does this happen?
Although the extremal black hole is well-known for its vanishing Hawking
temperature, and does not emit the usual late-time thermal radiation,
this tells us nothing about the transient behaviour right after
the gravitational collapse. In principle, it is possible that a
finite amount of quantum energy escapes before the state settle down
to the ground state, thus reducing the black hole mass.

To understand such a transient behaviour, one needs to calculate the
energy-momentum expectation values accurately everywhere, or equivalently
the relevant Bogolubov transformations valid for all retarded time,
not an easy task in general. But again with the help of Polyakov-Liouville
action, the problem becomes manageable for effective 2-D conformal modes.
The resulting semiclassical modification of the mass-to-charge ratio
was first demonstrated and estimated in reference \cite{Jaemo}, where $N$
chargeless Callan-Rubakov modes are quantized around extremal
Reissner-Nordstr\"{o}m black holes. Both analytic and numerical studies
revealed that the semiclassically corrected extremal black holes
obey the following mass-to-charge ratio, when $N\hbar/q^2$ is small;
\begin{equation}
\frac{\rm mass}{\rm charge}\:\simeq 1-\frac{N\hbar}{72\pi q^2} ,
\end{equation}
which is indeed consistent with the above interpretation of
one-loop WKB correction.

\vskip 5mm
The analogous shifts of the mass-to-charge ratio due to general
quantum fluctuations are yet to be calculated, but it is reasonable
to expect the same interpretation to hold for other cases as well.

\vskip 5mm
The main result (Eq. \ref{eq:remass}) states that to the first nonvanishing
order in $qB$, the one-loop corrected WKB exponent is simply the Schwinger
term but with quantum mechanically corrected mass rather than the
tree-level mass. The result is most sensible and assuring in that it is
exactly what one would expect for pair-production of ordinary charge
particles. But at the same time it is rather disappointing. Despite the
ever-increasing size of the bottom "cup" uniquely associated
with the near-extremal black hole pair-production, even the strongest
correction in the weak field limit can be explained away and does
not lead to new interesting physics.

Of course,
it is always a logical possibility that some nontrivial and large
physical effects are hidden in higher order terms, especially in $B$
independent one-loop contributions that must include a correction
to the Bekenstein-Hawking entropy.
Unlike our calculations here, unfortunately, the estimate of such
higher order terms in $qB$ appears a lot more sensitive to how
one treats the transitional "mouth" region, and thus is much more
difficult to carry out. For example, we have been rather cavalier
about the boundary at $r=r_B$ of the {\it truncated} black hole
and pretended that it is infinitely far away, which appears justifiable
only for the leading $qB$ behaviour found above.

One of the more problematic aspects of the Euclidean approach here
is  the fact that Euclidean path integral is ill-defined if dynamical
gravity is included, which we circumvented by substituting general
{\it matter} field for actual gravitational fluctuations. In other words,
the genuine operators governing small gravitational fluctuations are {\it
not} even elliptic, and in particular possess infinite number of
zero-eigenmodes, not to mention infinite number of negative eigenmodes.
Note that by substituting in general {\it matter} fluctuations as above,
we in effect concentrated on positive eigenmodes-modes only,

For these reasons, there is still
a long way to go before any concrete claim
can be put forward regarding the genuine one-loop WKB amplitude of this
process. One potentially promising extension of the present work is
to evaluate the analogous quantities in the context of string theory.
Note that what we have done here is essentially a calculation of
the partition function in single Euclidean black holes. By
generalizing to string one-loop partition functions, for
instance, we shall have the gravitational degree of freedom already
built-in. While there is no known exact conformal field theories
for 4-D nonextremal black holes, lower dimensional toy models are
available, such as a family of 3-D rotating
black holes \cite{BTZ}, described by orbifolds of the $SL(2,R)$
Wess-Zumino-Witten coset model \cite{Welch}. It might be possible to
gain further insight under such simplified settings.

\vskip  5mm
We are grateful to S. Giddings, K. Lee, J. Preskill, S. Trivedi, and
E. Weinberg for useful comments. We also thank Prof. C. Lee and the
organizers of the Sorak Symposium for the hospitality. This work is
supported in part by the US Department of Energy.

\end{document}